\documentclass[12pt]{article}

\usepackage{a4}
\usepackage{a4wide}

\usepackage{amsmath}
\usepackage{amscd}
\usepackage{amsfonts}
\usepackage{amssymb}
\usepackage{mathrsfs}
\usepackage{fancybox} 
\usepackage{enumerate}
\usepackage{cite}
\usepackage{bm}
\usepackage{amscd}
\usepackage{graphicx}
\usepackage[dvips]{color}
\usepackage{epsfig}

\makeatletter

\@addtoreset{equation}{section}
\makeatother
\if0
\usepackage{color}
\definecolor{blue1}{rgb}{0.5,0.15,0.10}
\definecolor{red1}{rgb}{0.8,0.1,0.10}
\usepackage[
debug,
dvipdfm,
colorlinks=true,
urlcolor=red1,
anchorcolor=red1,
citecolor=red1,
filecolor=red1,
linkcolor=red1,
menucolor=red1,
pagecolor=red1,
linktocpage=true,
pageanchor=false,
]{hyperref}    
\fi
\usepackage{amsthm} % This is for \newtheorem*
\newcommand{\wt}{\widetilde}

\def\={\stackrel{\bullet}{=}}

\def\({\left(}
\def\){\right)}
\def\[{\left[}
\def\]{\right]}

\def\mbf{\mathbf}

\def \be {\begin{equation}}
\def \ee {\end{equation}}
\def \beqa {\begin{eqnarray}}
\def \eeqa {\end{eqnarray}}
\def \beal#1 {\begin{align}#1\end{align}}
\def \bes#1 {\begin{equation}\begin{split}#1\end{split}\end{equation}}
\def \nn {\notag\\}

\makeatletter

\@addtoreset{equation}{section}
\makeatother

\begin{document}

\begin{titlepage}
\title{
\vspace{-2cm}
% \begin{flushright}
% \normalsize{ 
%  \\ 
% }
% \end{flushright}
\vspace{1.5cm}
Local Thermodynamics and Entropy for\\
Relativistic Hydrostatic Equilibrium
\vspace{.5cm}
}
\author{
Shuichi Yokoyama\thanks{syr18046[at]fc.ritsumei.ac.jp},\; 
\\[25pt] 
${}^{*}$ {\normalsize\it Department of Physical Sciences, College of Science and Engineering,} \\
{\normalsize\it Ritsumeikan University, Shiga 525-8577, Japan}
}

\date{}

\maketitle

\thispagestyle{empty}

% \vspace{.2cm}

\begin{abstract}
\vspace{0.3cm}
\normalsize

By refining the method proposed in arXiv:2010.07660, entropy current and entropy density for a relativistic hydrostatic equilibrium system with spherical symmetry are constructed as a non-N\"other conserved charge in the Einstein gravity with cosmological constant.
It is shown that the constructed entropy density satisfies both the local Euler relation and the first law of thermodynamics non-perturbatively with respect to the Newton constant. Finally the established relativistic thermodynamics is applied to one of the systems with uniform energy density as a crude model of a degenerate star and its local thermodynamic observables are determined analytically. 

\end{abstract}
\end{titlepage}
%\tableofcontents

\section{Introduction} 
\label{into}

Celestial bodies have attracted people from ancient time by not only their aesthetic appearances but also their physical properties. 
The fact that astronomical bodies are visible implies that they exist stably enough to be regarded as an equilibrium state, which is locally achieved by the balance between the attractive force of gravity and the repulsive force of matter inside.  
This suggests that astronomical bodies can be investigated as local thermodynamical objects described by local thermodynamical variables such as pressure and temperature whose dynamics obeys the laws of local thermodynamics.

The most innovative concept in thermodynamics is entropy \cite{clausius1867mechanical}. 
The laws of thermodynamics can be described most concisely by employing entropy. 
Then it is fundamental to ask {\it whether there exists entropy for stable astronomical bodies, and, if it exists, what its precise form is}.
Although it has been more than a century since the notion of entropy was created by Clausius \cite{clausius1867mechanical} and astronomical bodies were studied as thermodynamic objects by Helmholtz and Kelvin \cite{HelmholtzLXIVOT,kelvin1889popular}, a definite answer for this question seems to have remained unknown except for a specific case of constant energy density \cite{Oppenheim_2003}. 
The main reason for this will be that local thermodynamics including the correct definition of entropy density has not been established on curved spacetime.\footnote{ 
This does not mean the absence of definitions of (thermal) entropy density for relativistic field theory. However, the proposed definition in \cite{Aoki:2020nzm} is conceptually different from the conventional ones, for instance, \cite{Wald:1993nt,Gourgoulhon:2006bn,rezzolla2013relativistic,Andersson:2020phh}, as described below. Another approach is to determine entropy density so as to satisfy standard thermodynamic relations locally \cite{Oppenheim:2001nx,Oppenheim_2003}. 
In order to justify this approach, the assumed thermodynamic relations have to be verified finally, though the proof has not been found or complete because the correct local form of the specific volume element has not been specified, as explained below.
} 
However, this question is very basic and must be answered because the existence of entropy provides the ground to apply basic results derived by assuming statistical ensemble to local thermodynamic variables as seen in classic arguments for the understanding of physics of astronomical bodies. 

Recently the author and collaborators have proposed a method to construct entropy current and entropy density as a conserved current and a conserved charge density, respectively, for general field theory defined on general curved spacetime with covariantly conserved energy momentum tensor even without any global symmetry \cite{Aoki:2020nzm}. (See also \cite{Aoki:2020prb}.)
In this paper, the author aims at answering the above question adopting the proposed method by simplifying a situation in such a way to approximate an astronomical body as a relativistic hydrostatic equilibrium system with spherical symmetry.
Then the question is positively answered and the expression of the entropy density endowed with laws of local thermodynamics is explicitly determined for an arbitrary value of the Newton constant as shown below. 

\section{Relativistic hydrostatic equilibrium system}
\label{HydrostaticSyetem}

Consider a $d$-dimensional system of fluid of spherically symmetric configuration in hydrostatic equilibrium \cite{PhysRev.35.904,PhysRev.55.374}.
Such a fluid may be described by a so-called perfect fluid, whose energy stress tensor is given by
\be 
T_{\mu\nu}= (\rho+p) u_\mu u_\nu + p g_{\mu\nu} ,
\label{PerfectFluid}
\ee
where $\rho$ is the total energy density, $p$ is the pressure, $u^\mu$ is the fluid velocity normalized by $g_{\mu\nu}u^\mu u^\nu=-1$, $g_{\mu\nu}$ is a metric tensor in a general rotationally symmetric form  
\beal{
g_{\mu\nu}\mathrm dx^\mu \mathrm dx^\nu
=& - f \mathrm dt^2 +h \mathrm dr^2 + r^2\tilde g_{ij}\mathrm dx^i\mathrm dx^j
\label{metric}
}
where $\tilde g_{ij}$ is the metric for the $d-2$ dimensional internal space whose the Ricci curvature tensor is $\tilde{\rm Ric}_{ij}=(d-3)\tilde g_{ij}$, such as the unit sphere.
The rotational symmetry constrains all the functions $u^\mu, \rho, p, f, h$ to depend only on the radial coordinate $r$.
 
For the purpose to understanding sufficiently low energy physics of the above hydrostatic equilibrium system, it is sufficient to consider gravitational interaction described by the Einstein-Hilbert action. Then the hydrostatic equilibrium is described by the Einstein equation. 
In order for fluid to couple to the Einstein gravity consistently, its energy stress tensor is required to satisfy the covariantly conservation equation in the gravitational background, $\nabla_\mu T^\mu\!_\nu=0$. 
For the perfect fluid \eqref{PerfectFluid}, the covariant conservation equation leads to the following relativistic fluid equations
\beal{
\check\nabla \rho + (\rho + p) \vartheta =& 0 , 
\label{CCEt} \\
(\rho + p )\check \nabla u_\mu + \bar\nabla_\mu p  =&0, 
\label{CCEr}
}
where $\check \nabla := u^\mu \nabla_\mu, \bar\nabla_\mu := \nabla_\mu +u_\mu\check\nabla$ and $\vartheta=\bar\nabla_\mu u^\mu $ is the so-called expansion of the fluid. 
For later use, the fluid equations are rewritten in an observing frame with the fluid radially moving, $(u^\mu)=(u^t,u^r,\vec0)$, though the system is finally analyzed by taking the limit of the comoving frame of the fluid velocity $(u^\mu)=(1/\sqrt f,0,\vec0)$. 
In this radially moving frame, the normalization condition is 
\be 
-f(u^t)^2+h(u^r)^2 =-1,
\label{Normalization}
\ee
nad the relativistic fluid equations \eqref{CCEt}, \eqref{CCEr} reduce to  
\beal{
\rho' + (\rho + p) (\log (\sqrt{|g|}u^r))' = 0 , 
\label{CCEt2} \\
p' +(\rho + p )(\log (fu^t))'  =0 , 
\label{CCEr2}
}
where the prime means the differentiation with respect to the radial coordinate. 
Note that \eqref{CCEt2} is obtained from \eqref{CCEt}, while the radial component of \eqref{CCEr} reduces to \eqref{CCEr2} by using \eqref{Normalization}. 

In the comoving frame of no energy motion, $u^r\to0$, the condition of the covariant conservation \eqref{CCEr2} reduces to 
\be 
p'
= -\frac{(p+\rho)}{2}(\log f)', 
\label{CCEr3}
\ee
and the Einstein equation 
\beal{
(\log h)'=\frac{2 r h }{d-2}(8\pi G_N\rho +\Lambda) -\frac{(d-3) \left(h-1\right)}{r},
\label{EOM1} \\
(\log f)'=\frac{2 r h }{d-2}(8\pi G_N p -\Lambda)+ \frac{(d-3) \left(h-1\right)}{r},
\label{EOM2}
}
where $G_N$ is the Newton constant and $\Lambda$ is the cosmological constant.
Introducing a new variable $M_r$ defined by 
\be 
M_r :={ r^{d-3} \over 2 G_N } \(1 - {1\over h}  - {2\Lambda r^2 \over (d-2)(d-1) } \),
\label{M_r}
\ee
one can rewrite \eqref{EOM1} as 
\beal{ 
\rho=& \frac{(d-2) M_r'}{8\pi r^{d-2} },
\label{rho}
} 
while \eqref{EOM2} as 
\beal{
(\log f)'=& \frac{ r^{d-1}  \frac{1} {(d-2)} (8\pi G_N p -\Lambda) + (d-3) G_NM_r }{r^{d-2} - 2r G_NM_r}.  
\label{EOM2'}
}
Plugging this into \eqref{CCEr3} with $d=4$ and $\Lambda=0$, one obtains the Tolman-Oppenheimer-Volkov (TOV) equation given in \cite{PhysRev.55.374}.  

\section{Local thermodynamic relations and entropy}
\label{Entropy}

As mentioned in introduction, any equilibrium system will be expected to be described by a set of thermodynamic variables obeying thermodynamic relations. Then how does one find the entropy for a hydrostatic equilibrium system in curved spacetime as described above?
The author answers this question adopting the method proposed in \cite{Aoki:2020nzm}, which provides a general prescription to construct a conserved quantity for a closed system with its energy momentum tensor covariantly conserved and even without any global symmetry. 
The prescription is to find a vector field to satisfy what is called a conservation equation $T^\mu\!_\nu \nabla_\mu \xi^\nu = 0$. 
Once such a vector field $\xi^\nu$ is found, an associated conserved current is constructed as $J^\mu[\xi] = \sqrt{|g|} T^\mu\!_\nu \xi^\nu$ and a conserved charge $Q[\xi]=\int d^{d-1}x J^t[\xi]$. 
In order to construct entropy current associated with fluid, an associated vector field is assumed to be proportional to its fluid velocity, $\xi^\nu =-\zeta u^\nu,$ where $\zeta$ is a scalar function, so that the ansatz respects general covariance.%
\footnote{ 
Since any physical fluid has the unique flow velocity, this prescription is easily available to any single fluid. Even for a case with fluid multi-constituent, the prescription is available by setting the ansatz of the vector field to be in a form of a linear combination of the flow velocity of each constituent. 
}
Substituting this ansatz into the conservation equation leads to $-\rho \check\partial \zeta + \zeta p \vartheta = 0.$
In a radially moving frame, this becomes $u^r\zeta' = {p \vartheta \over \rho }\zeta$, where $\zeta$ is assumed to depend only on $r$ for a spherically symmetric equilibrium system. In this frame, the expansion is computed as $\vartheta=u^r \frac{ - \rho' }{\rho + p} $, 
where \eqref{CCEt2} was used. 
Substituting this back, one obtains%
\footnote{ 
This equation holds even in the limit of the rest frame $u^r\to0$, because it holds in the general radially moving frame with the metric tensor satisfying the Einstein equation in the frame. Such a solution for the Einstein equation with radial steady flow is assured at least locally by a general theory of differential equation by adding terms suitably, and it is sufficient for obtaining the differential equation \eqref{zeta}. 
}
\be 
\zeta' = -{p \over \rho }\frac{ \rho' }{\rho + p}\zeta. 
\label{zeta}
\ee
This determines $\zeta$ in terms of macroscopic observables. 
Then the conserved current associated with the vector field is computed as 
\beal{
s^\mu
=\sqrt{|g|} \rho \zeta u^\mu. 
\label{EntropyCurrent}
}
This is constructed to satisfy the continuity equation $\partial_\mu s^\mu=0$. 
This can be confirmed as follows. 
In the spherically symmetric equilibrium system, $\partial_\mu s^\mu=\partial_r s^r$. On the other hand,
\beal{
\partial_r s^r
=& \sqrt{|g|}u^r  ( (\log(\sqrt{|g|}u^r))' \rho \zeta + \rho' \zeta + \rho \zeta' ) 
\nn 
=& \sqrt{|g|}u^r  ( \frac{ - \rho' }{\rho + p}  \rho \zeta + \rho' \zeta -{p }\frac{ \rho' }{\rho + p}\zeta) 
=0,
}
where \eqref{CCEt2} and \eqref{zeta} were used. This completes the confirmation. 

As a result, $s^\mu$ is a conserved current and its time component $s^t=:s$ is a conserved charge density. 
These conserved quantities are claimed to be the entropy current and the entropy density of the system, respectively \cite{Aoki:2020nzm}.%
\footnote{ 
It is worth commenting that the relation between entropy density and entropy current presented in the paper is different from the conventional one seen in some textbooks and reviews of relativistic fluid dynamics, for instance \cite{Gourgoulhon:2006bn,rezzolla2013relativistic,Andersson:2020phh}, as $\tilde s^\mu=\tilde s u^\mu$, where $\tilde s$ is their entropy density measured in the rest frame of the fluid, on general curved spacetime. The two entropy densities are conceptually different in the regard that the conventional one is defined as a scalar while the proposed one is as a charge density. This conceptual difference leads to that in their transformation properties and thus gives rise to a significant one to define the entropy density on curved spacetime from that in the freely falling observing frame.
Indeed, they clearly become different functions unless the time component of the flow velocity is trivial. Therefore, if the conventional definition was adopted as the entropy density, then it would not be possible to derive the local thermodynamic relations presented in this paper. 
The author would like to thank a referee for asking a question on this point. 
}
In order to provide strong evidence, let me show below that this identification leads to the local Euler's relation and the first law of thermodynamics concurrently in the comoving frame, which are respectively described by
\beal{
T s =& u +p v, 
\label{Euler} \\
T ds =& du+ pdv, 
\label{1stLaw} 
}
where $v$ is the measure of the volume element at a constant time slice, $v =r^{d-2} \sqrt{h\tilde g}$, and $u$ is the internal energy density, $u=\rho v$, and $T$ is a scalar corresponding to the temperature defined to satisfy both \eqref{Euler} and \eqref{1stLaw}. 
To show this, first solve \eqref{zeta} as 
\be 
\zeta =\beta_0 u^t f (1 + \frac p\rho) , 
\label{zetasol}
\ee 
where $\beta_0$ is an integration constant. 
Indeed, assuming \eqref{zetasol} one can calculate $
\zeta'
= \zeta \partial_r\log\( (u^t f) (1 + \frac p\rho)\right) 
= \zeta ( - \frac{p\rho'}{\rho(\rho +p)} ), $
where \eqref{CCEr2} was used. This satisfies \eqref{zeta}. 
Plugging \eqref{zetasol} into \eqref{EntropyCurrent}, $s^\mu$ can be written as $s^\mu= \beta_0 u^t f^{\frac32} (u + pv) u^\mu$. 
Therefore the Euler relation \eqref{Euler} holds if and only if the scalar $T$ or its inverse $\beta:=1/T$ is given by 
$\beta =\beta_0 (u^t)^2 f^{\frac32} . $
In the comoving frame, $u^t \to 1/\sqrt f$, this becomes 
\be 
\beta =\beta_0 \sqrt f. 
\label{betaComoving}
\ee
In what follows, the frame is fixed by the comoving one. Then 
the derivative of $\beta$ with respect to $r$ is computed as 
\be 
\beta'
= -\frac{p'}{p+\rho}\beta, 
\label{TemperaturePDE}
\ee 
where \eqref{CCEr3} was used, and thus that of $s=s^t$ is $
s' = \beta'(u+vp) + \beta(u+vp)'=\beta( u'+pv').$
This is the first law of thermodynamics \eqref{1stLaw}. This completes the proof. As a result, it concludes that $s^\mu, s, \beta$ are interpreted as the entropy current, the entropy density, and the inverse temperature, respectively.

Comments are in order. 
The (inverse) temperature introduced to satisfy the local Euler relation and the first law of thermodynamics in \eqref{betaComoving} exactly matches the one derived by Tolman as the 'proper temperature' for a local observer \cite{PhysRev.35.904}. 
This temperature was also derived by different ways in standard textbooks \cite{alma991013077249704706,Misner:1974qy}.
% p.263~\cite{alma991013077249704706}, p.568~\cite{Misner:1974qy}.
On the other hand, the 'entropy vector' and 'proper entropy density' were introduced to discuss hydrostatic equilibrium in \cite{1935ApJ....82..435H} without explicit expression, while the entropy density and the entropy current are explicitly determined in this paper. 
From the entropy density, one can compute the local entropy inside the sphere of radius $\bar r$ denoted by $\mbf S_{\bar r}$ as $S_{\bar r} = \int_{\mbf S_{\bar r}} \mathrm d^{d-1}x s= \int_0^{\bar r} dr (\rho+p)V_r'/T$, where $V_{\bar r}=\int_{\mbf S_{\bar r}} \mathrm d^{d-1}x v$ is the volume of the spherical region within the radius $r$.
These results were obtained without restricting the value of the Newton constant. Therefore the thermodynamic quantities and thermodynamic relations hold non-pertubatively with respect to the Newton constant. 
Finally, in \cite{Oppenheim_2003}, the entropy density with respect to a non-proper coordinates $s_0$ is introduced to satisfy another local thermodynamic relation in the non-proper coordinates such that $s_{0} = \beta(\rho+p) -\mu n$, where the chemical potential $\mu$ is also introduced, and it is claimed that multiplying this by 'a tiny volume element $V$' leads to the thermodynamic relations such as \eqref{Euler} and \eqref{1stLaw} including the term of chemical potential.\footnote{ 
Once the local Euler's relation \eqref{Euler} and the first law of thermodynamics \eqref{1stLaw} are proved, it is not difficult to include the terms of chemical potential and number density of particles formally by shifting the definition of entropy density as $s=\check s + \sum_i\frac{\mu_i}T n_i$ so as to extend the local Euler's relation to $T\check s = u +p v -\sum_i \mu_i n_i$. In order to satisfy the first law of thermodynamics $Td\check s = du +p dv -\sum_i \mu_i dn_i$, it is required to satisfy $\sum_i d(\mu_i/T) n_i = 0$. For the case of one species of particle, the condition reduces to $d(\mu/T) = 0$, or, $\mu\propto T$, which agrees with the result in \cite{Oppenheim:2001nx,Oppenheim_2003,glendenning2012compact}.
} However, the 'tiny volume element $V$' is not specified in the main text of \cite{Oppenheim_2003}, while the quantity denoted by $V$ is introduced in Appendix to compute the entropy of the system with constant energy density to satisfy $dV=4\pi r^2 \sqrt{g_{rr}} dr$. If this definition of $V$ in Appendix of \cite{Oppenheim_2003} is adopted in the main text of \cite{Oppenheim_2003}, then the thermodynamic relations such as \eqref{Euler} and \eqref{1stLaw} do not hold. In this context, a new claim of this paper can be rephrased to say that the local specific volume element (called 'a tiny volume element $V$' in \cite{Oppenheim_2003}) has to be given as the measure of the volume element at a constant time denoted by $v$ in order to prove the local thermodynamic relations.\footnote{ 
This specification of the correct local form of the specific volume element is not trivial at all because such a tiny volume element is given to be proportional to the volume per particle $1/n$ or the inverse density $1/\rho$ in the textbooks of fluid dynamics and general relativistic hydrodynamics the author has ever seen, for instance, \cite{Weinberg:1972kfs,Misner:1974qy,rezzolla2013relativistic,zel2014stars}.
}

\section{Application to a spherical star with uniform density} 
\label{Application} 

In order to see how the relativistic thermodynamics works, it is below applied to a perfect fluid with spherical configuration of radius $R$ and uniform energy density in the dark energy specified by the cosmological constant $\Lambda$ \cite{Misner:1974qy,glendenning2012compact,Oppenheim_2003}.
This could be a crude model of a degenerate core deep inside of a massive star and a compact star of degenerate matter \cite{10.1093/mnras/87.2.114}. 
The analysis is performed with the general dimension $d\geq3$ and assumes that the cosmological constant is sufficiently small.

Inside the star or its core, $r\leq R$, the energy density is constant, $\rho=\rho_0$, so that $M_r$ can be determined from \eqref{rho} as 
\beal{
M_r 
={8\pi \rho_0 r^{d-1}\over (d-2)(d-1)} 
=:M(r) 
\label{M(r)}
}
where the integration constant was determined for $M_r$ to vanishes at the center. 
Then from \eqref{M_r}, $h =\frac{r_0^2}{r_0^2-r^2} $, 
where $r_0 =\sqrt \frac{(d-2) (d-1)}{ 2 (8\pi G_N \rho_0 +\Lambda) }$, and from \eqref{EOM2'} 
\beal{
(\log f)'  = \frac{ (d-3) 8\pi G_N \rho_0 r + (d-1) r(8\pi G_N p - \Lambda)}{(d-2)(d-1) - 16\pi G_N \rho_0 r^2} .
}
Plugging this into \eqref{CCEr3}, one can solve the TOV equation as
\be  
p=\frac{-C \sqrt{r_0^2-r^2}+d-3 -2 \wt\Lambda }{C \sqrt{r_0^2-r^2}-(d-1) }\rho_0, 
\ee
where $C$ is an integration constant and $\wt\Lambda:=\Lambda/(8\pi G_N\rho_0)$.  
Fixing the integration constant so as for the pressure to vanish at the surface of the star, one can determine the radial dependence of pressure as
\beal{
p=&\frac{ \left(\sqrt{r_0^2-r^2}-\sqrt{r_0^2-R^2}\right) \left(d-3  -2 \wt\Lambda \right)}{- \left(d-3  -2 \wt\Lambda \right) \sqrt{r_0^2-r^2}+(d-1)\sqrt{r_0^2-R^2}  } \rho_0.
\label{psol}
}
Note that in order for the configuration to be stable, the pressure has to be positive and finite, which is satisfied if and only if $d-3  -2 \wt\Lambda > 0 , ~ - \left(d-3  -2 \wt\Lambda \right) r_0+(d-1)\sqrt{r_0^2-R^2} > 0 $. 
This condition is rewritten as follows.  
\be 
R
< \sqrt{ \frac{( d-2 -\wt\Lambda)(d-2) }{ (d-1) 4\pi G_N \rho_0 }}=:R_{cr},
~~~ {\rm max}\{ \frac{\Lambda}{(d-3)4\pi G_N} , \frac{-\Lambda}{8\pi G_N} \} <  \rho_0. 
\label{Condition}
\ee
% The results up to here are already known. 
% A new result in the paper is to make it possible to also determine the temperature of the system analytically by solving \eqref{TemperaturePDE} as 
The temperature is also determined analytically by solving \eqref{TemperaturePDE} as%
\footnote{  This result agrees with the result (D7) in \cite{Oppenheim_2003} up to the initial conditions, whose relation is $T(R)=T_0/k(R)$.  }
\be 
T 
=T(R) { p+\rho_0 \over \rho_0 } 
= T(R) \frac{ 2 \sqrt{r_0^2-R^2} (1+ \wt\Lambda )}{- (d-3  -2 \wt\Lambda ) \sqrt{r_0^2-r^2}+(d-1)\sqrt{r_0^2-R^2}  }  , 
\label{Tsol}
\ee
where an integration constant appears as the surface temperature of the star $T(R)$. This is fixed by input of observational data.
In particular, the temperature at the center is given by the one at the surface as 
\be 
T(0) = \frac{ 2(1+ \wt\Lambda )\sqrt{1-(R/r_0)^2}}{- (d-3  -2 \wt\Lambda ) +(d-1)\sqrt{1-(R/r_0)^2}  }T(R). 
\ee
Note that the core maximum temperature becomes higher and higher as the radius approaches the critical value $R_{cr}$. 
Substituting \eqref{psol} and \eqref{Tsol} into \eqref{Euler} one obtains the entropy density as $s = \frac {u}{T(R)}$, where $u=\frac{\rho_0r_0}{\sqrt{r_0^2 -r^2}} \sqrt{\tilde g}$. 
Interestingly, this system has another form of the local Euler's relation with global (surface) temperature, which shows that while the local temperature monotonically decreases as it approaches the surface away from the core, the internal energy density and the entropy one monotonically increase for a uniform stellar object in hydrostatic equilibrium.\footnote{
This phenomenon of the entropy increase to the surface was already observed and its connection to the area law of the black hole entropy  \cite{Bekenstein:1972tm,Bekenstein:1973ur,Bardeen:1973gs} was insightfully pointed out in \cite{Oppenheim_2003}. 
The author would like to thank a referee for pointing this out. 
} 

Outside the star, $r\geq R$, there is no fluid. Thus the density and pressure of fluid vanish, so does the entropy density from \eqref{Euler}. Therefore, from \eqref{rho}, $M_r$ has to be a constant, which is determined as $M_r=M(R)$ from the continuity condition at the surface of the star. 
On the other hand, from \eqref{EOM1} and \eqref{EOM2}, $(\log f)' = -(\log h)'$, so that $f$ is proportional to $1/h$. 
Setting the boundary condition at infinity for the metric to be asymptotic to the standard (anti-) de Sitter space metric, one can fix the proportional coefficient as $1$, so that $f,h$ are determined as $f=1/h={-2\Lambda r^2 \over (d-2)(d-1) } + 1 - {2G_N M(R) \over r^{d-3}}=:f(r)$. Then the metric \eqref{metric} in the matter-empty region is given by the Kottler metric with the so-called gravitational mass $M(R)$.
In the matter-empty region the thermodynamic relations \eqref{Euler} and \eqref{1stLaw} are trivial, though the temperature is non-trivially determined from \eqref{betaComoving} as $T=T(R) \sqrt{f(R)/f(r)} $. 
It can be seen that the local temperature is formally divergent at the location of the horizon of the black hole whose mass is identical to the stellar gravitational mass. 
Note that in order for $f(r)$ to be always positive, the radius of the star is required to be greater than that of any event horizon of the Schwarzschild black hole with the same mass $M(R)$. For $\Lambda=0, d=4$ this condition is in fact satisfied by the above constraint \eqref{Condition}, which reduces to $M(R)<4R/(9G_N)$. 

It is not difficult to compute local additive quantities inside the sphere of radius $r$ with $r\leq R$. 
The local internal energy $U_r$ is computed as $U_r = \rho_0 V_r$, where 
$V_r =\tilde V \int_0^r \mathrm dr' r'^{d-2} { r_0 \over \sqrt{r_0^2 -r'^2} } $. Particularly if one chooses $d=4$ and the internal manifold is the unit sphere, then
$V_r= 2\pi r_0 \left(r_0^2 \arctan \frac{r}{\sqrt{r_0^2-r^2}}-r \sqrt{r_0^2-r^2}\right)$. 
The local entropy is $S_r= \int_{\mbf S_{r}} \mathrm d^{d-1}x s = \frac {U_r}{T(R)}$. 
Amusingly, from these results, the extensive behavior is easily seen for the set of local quantities, $(S_r,U_r,V_r)$, which is not simply by integrating both sides of the local Euler's relation \eqref{Euler}, since $T$ and $p$ are non-trivial functions of the radial coordinate.
As a result the total entropy of the system is obtained as $S_R = \frac {U_R}{T(R)}$, which matches the result (D15) in \cite{Oppenheim_2003} with $n_0=\mu_0=0$.\footnote{ 
There seems to be a few minor errors in (D15) in \cite{Oppenheim_2003}. One is '$\mu$' in the right-hand side of the upper equation, which must be $\mu_0$ as easily confirmed from (D12). Another is '$\mu$' in the lower equation, which will be replaced by $\mu\to\frac{3\mu_0}{2R}$. After the replacement, a constant term $N\mu_0/(2T_0)$ may be added to (D15) for the correct result. 
}

\section{Discussion} 

Entropy current and entropy density for a relativistically hydrostatic equilibrium system with spherical symmetry have been constructed as a conserved non-N\"other charge by improving the method in \cite{Aoki:2020nzm}. The refined prescription presented in the paper is to choose a vector field for the construction of a conserved charge so as to be proportional to the fluid velocity of the system. This new prescription allows one to construct the entropy density of the system uniquely for any physical fluid and to use the relativistic fluid equation for the construction.
It has been shown that the entropy density concurrently satisfies the local Euler relation and the first law of thermodynamics with the local temperature identical to the one determined by Tolman \cite{PhysRev.35.904} non-perturbatively with the Newton gravitational constant.
After the foundation of relativistic thermodynamics has been established, 
it has been applied to a spherically symmetric hydrostatic system of a perfect fluid with uniform energy density as a crude model of a degenerate core deep inside a massive star and a compact star. It has been shown that local thermodynamic observables are precisely determined even deep inside the core in this model. 

An important virtue of the presented method is that this is applicable  not only to an equilibrium system but also to a non-equilibrium one. In particular, the constructed entropy density in this paper is ensured to be the one left after an evolutionary dynamical system reaches the equilibrium with spherical symmetry. It would be very interesting to apply the presented method to a non-equilibrium system. 

Further applications of the established relativistic thermodynamics to a more realistic situation and to the construction of a model of celestial bodies will be reported in the near future. 

\section*{Acknowledgement}
This work is supported in part by the Grant-in-Aid of the Japanese Ministry of Education, Sciences and Technology, Sports and Culture (MEXT) for Scientific Research (No.~JP22K03596). 

\bibliographystyle{utphys}
\bibliography{stellar}

\providecommand{\href}[2]{#2}\begingroup\raggedright\begin{thebibliography}{10}

\bibitem{clausius1867mechanical}
R.~Clausius, T.~Hirst, and J.~Tyndall, {\em The Mechanical Theory of Heat: With
  Its Applications to the Steam-engine and to the Physical Properties of
  Bodies}.
\newblock J. Van Voorst, 1867.

\bibitem{HelmholtzLXIVOT}
H.~L.~F. von Helmholtz, ``Lxiv. on the interaction of natural forces,'' {\em
  Philosophical Magazine Series 1} {\bf 11} 489--518.

\bibitem{kelvin1889popular}
W.~Kelvin, {\em Popular Lectures and Addresses}.
\newblock Nature series. Macmillan and Company, 1889.

\bibitem{Oppenheim_2003}
J.~Oppenheim, ``Thermodynamics with long-range interactions: From ising models
  to black holes,'' {\em Physical Review E} {\bf 68} (July, 2003).

\bibitem{Aoki:2020nzm}
S.~Aoki, T.~Onogi, and S.~Yokoyama, ``{Charge conservation, entropy current and
  gravitation},'' {\em Int. J. Mod. Phys. A} {\bf 36} (2021), no.~29, 2150201,
  \href{http://arXiv.org/abs/2010.07660}{{\tt 2010.07660}}.

\bibitem{Wald:1993nt}
R.~M. Wald, ``{Black hole entropy is the Noether charge},'' {\em Phys. Rev. D}
  {\bf 48} (1993), no.~8, 3427--3431,
  \href{http://arXiv.org/abs/gr-qc/9307038}{{\tt gr-qc/9307038}}.

\bibitem{Gourgoulhon:2006bn}
E.~Gourgoulhon, ``{An Introduction to relativistic hydrodynamics},'' {\em EAS
  Publ. Ser.} {\bf 21} (2006) 43--79,
  \href{http://arXiv.org/abs/gr-qc/0603009}{{\tt gr-qc/0603009}}.

\bibitem{rezzolla2013relativistic}
L.~Rezzolla and O.~Zanotti, {\em Relativistic Hydrodynamics}.
\newblock EBSCO ebook academic collection. OUP Oxford, 2013.

\bibitem{Andersson:2020phh}
N.~Andersson and G.~L. Comer, ``{Relativistic fluid dynamics: physics for many
  different scales},'' {\em Living Rev. Rel.} {\bf 24} (2021), no.~1, 3,
  \href{http://arXiv.org/abs/2008.12069}{{\tt 2008.12069}}.

\bibitem{Oppenheim:2001nx}
J.~Oppenheim, ``{A Thermodynamic sector of quantum gravity},''
  \href{http://arXiv.org/abs/gr-qc/0112001}{{\tt gr-qc/0112001}}.

\bibitem{Aoki:2020prb}
S.~Aoki, T.~Onogi, and S.~Yokoyama, ``{Conserved charges in general
  relativity},'' {\em Int. J. Mod. Phys. A} {\bf 36} (2021), no.~10, 2150098,
  \href{http://arXiv.org/abs/2005.13233}{{\tt 2005.13233}}.

\bibitem{PhysRev.35.904}
R.~C. Tolman, ``On the weight of heat and thermal equilibrium in general
  relativity,'' {\em Phys. Rev.} {\bf 35} (Apr, 1930) 904--924.

\bibitem{PhysRev.55.374}
J.~R. Oppenheimer and G.~M. Volkoff, ``On massive neutron cores,'' {\em Phys.
  Rev.} {\bf 55} (Feb, 1939) 374--381.

\bibitem{alma991013077249704706}
I.~B. I.~B. Zel'dovich, {\em Relativistic astrophysics}.
\newblock University of Chicago Press, Chicago, 1971 - 1983.

\bibitem{Misner:1974qy}
C.~W. Misner, K.~S. Thorne, and J.~A. Wheeler, {\em {Gravitation}}.
\newblock W. H. Freeman, San Francisco, 1973.

\bibitem{1935ApJ....82..435H}
O.~{Heckmann}, ``{REVIEW: Relativity, Thermodynamics and Cosmology, by R. C.
  Tolman},'' {\em apj} {\bf 82} (Dec., 1935) 435.

\bibitem{glendenning2012compact}
N.~Glendenning, {\em Compact Stars: Nuclear Physics, Particle Physics and
  General Relativity}.
\newblock Astronomy and Astrophysics Library. Springer New York, 2012.

\bibitem{Weinberg:1972kfs}
S.~Weinberg, {\em {Gravitation and Cosmology}: {Principles and Applications of
  the General Theory of Relativity}}.
\newblock John Wiley and Sons, New York, 1972.

\bibitem{zel2014stars}
Y.~Zel'dovich and I.~Novikov, {\em Stars and Relativity}.
\newblock Dover Books on Physics. Dover Publications, 2014.

\bibitem{10.1093/mnras/87.2.114}
R.~H. Fowler, ``{On Dense Matter},'' {\em Monthly Notices of the Royal
  Astronomical Society} {\bf 87} (12, 1926) 114--122,
  \href{http://arXiv.org/abs/https://academic.oup.com/mnras/article-pdf/87/2/114/3623303/mnras87-0114.pdf}{{\tt
  https://academic.oup.com/mnras/article-pdf/87/2/114/3623303/mnras87-0114.pdf}}.

\bibitem{Bekenstein:1972tm}
J.~Bekenstein, ``{Black holes and the second law},'' {\em Lett. Nuovo Cim.}
  {\bf 4} (1972) 737--740.

\bibitem{Bekenstein:1973ur}
J.~D. Bekenstein, ``{Black holes and entropy},'' {\em Phys. Rev. D} {\bf 7}
  (1973) 2333--2346.

\bibitem{Bardeen:1973gs}
J.~M. Bardeen, B.~Carter, and S.~Hawking, ``{The Four laws of black hole
  mechanics},'' {\em Commun. Math. Phys.} {\bf 31} (1973) 161--170.

\end{thebibliography}\endgroup

\end{document}